\documentclass[]{aipproc}
\layoutstyle{6x9}

\def\micron{\mu m}
\def\ev{\hbox{eV}}
\def\kev{\hbox{keV}}
\def\gev{\hbox{GeV}}
\def\GeV{\gev}
\def\tev{\hbox{TeV}}
\def\Ecut{{\cal E}_{cut}}
\def\Ephot{{\cal E}_\gamma}
\def\ebgphot{{\epsilon}_\gamma}
\def\me{m_e}
\def\nature{{\it Nature}}
\def\lesssim{\mathrel{\hbox{\rlap{\hbox{\lower4pt\hbox{$\sim$}}}\hbox{$<$}}}}
\def\gtrsim{\mathrel{\hbox{\rlap{\hbox{\lower4pt\hbox{$\sim$}}}\hbox{$>$}}}}
\let\la=\lesssim            
\let\ga=\gtrsim 

\begin{document}

\title[High Redshift Gamma Ray Bursts]{Gamma Ray Bursts as Probes of
 the First Stars}
\author{James E. Rhoads}{
  address={STScI, 3700 San Martin Dr., Baltimore, MD 21210, USA},
  email={rhoads@stsci.edu},
}

\begin{abstract}
The redshift where the first stars formed is an important and unknown
milestone in cosmological structure formation. The evidence linking
gamma ray bursts (GRBs) with star formation activity implies that the
first GRBs occurred shortly after the first stars formed.  Gamma ray
bursts and their afterglows may thus offer a unique probe of this
epoch, because they are bright from gamma ray to radio wavelengths and
should be observable to very high redshift.  Indeed, our ongoing
near-IR followup programs already have the potential to detect bursts
at redshift $z \sim 10$.  In these proceedings, we discuss two
distinct ways of using GRBs to probe the earliest star formation.
First, direct GRB counts may be used as a proxy for star formation
rate measurements.  Second, high energy cutoffs in the GeV spectra of
gamma ray bursts due to pair production with high redshift optical and
ultraviolet background photons contain information on early star
formation history.  The second method is observationally more
demanding, but also more rewarding, because each observed pair
creation cutoff in a high redshift GRB spectrum will tell us about the
integrated star formation history prior to the GRB redshift.
\end{abstract}

\maketitle

\section{Introduction}
The high redshift frontier of observational cosmology currently stands at
redshifts $z \approx 6$.  The current redshift record is a quasar at
$z=5.8$, and a few galaxies are known at marginally lower redshift.
Beyond $z=6$, we have yet to identify any individual objects.  We do
know that hydrogen was predominantly neutral at redshifts $z \ga 30$
based on the observed anisotropy of the cosmic microwave background,
which would be smoothed out by Thomson scattering if the free
electron density at $z \ga 30$ were too great.  The
redshift range $6 \la z \la 30$ remains unknown territory.  It is a very
interesting territory, too, for it should include the formation of the
first stars, galaxies, and quasars, and certainly includes the epoch 
at which hydrogen was reionized. 

Searches for starlight (and other rest-frame near ultraviolet tracers)
can make incremental progress into the low-redshift end of this
period.  However, these methods face a practical limit where the Lyman
break redshifts out of the optical window to the near-infrared, at $z
\approx 7$.  At higher redshift, essentially no flux is expected in the optical
window (observed wavelengths $0.36 \micron \la \lambda_{obs} \la 1
\micron$).  Atmospheric conditions and present detector technologies conspire
to make searches at $\lambda_{obs} \ga 1 \micron$ much less efficient.
Future instrumentation like the Next Generation Space Telescope (NGST)
promise extensions of ``conventional'' optical methods to the observed
near-IR and thus to redshifts $z \gg 6$, but this may be a decade or
more away.  In the meantime, we expect the upcoming extension of our
Large Area Lyman Alpha (LALA) survey (Rhoads et al 2000; Malhotra et
al 2001) to $z=6.6$ to be at or near the practical limit for some
years.

We would like to find tracers of $z>6$ objects that are accessible
now.  Fortunately, this is possible so long as we are willing
to use something besides starlight.  In practice, this means
higher energy photons ($\gamma$ and x-rays), since lower energies
still face either confusion or sensitivity issues.

Gamma ray bursts (GRBs) are an excellent candidate for detection at high
redshift because the bursts and their afterglows are extremely bright
at all wavelengths.  Two conditions must be met for such a candidate
to work well.  First, there should be a reasonable expectation that the
object exists at high redshift; and second, it should be detectable there.

The best argument that gamma ray bursts should occur at high redshift
comes from the growing body of evidence linking GRBs to star formation
activity (and hence presumably to the deaths of massive, short-lived
stars): GRB host galaxy colors are
characteristically blue (Fruchter et al 1999); the spatial
distribution of GRBs on their hosts matches expectations for
hypernova models (Bloom, Kulkarni, \& Djorgovski 2000); and the
emission lines of GRB host galaxies are unusually strong
(Fruchter et al 2001).  Structure formation models yield estimated
redshifts $z \sim 15 \pm 5$ for the first stars to form in the
universe (cf. Barkana \& Loeb 2001).  This is supported by studies of
heavy element abundances: It has proven extremely difficult to find
objects with primordial (i.e., big bang nucleosynthesis) abundances at
any redshift currently accessible.  The immediate inference is that a
substantial generation of stars must have existed at earlier redshifts
to produced the ubiquitous metals.  The association of GRBs with star
formation then implies that the first GRBs also occurred in the
redshift range $z \sim 15 \pm 5$.  

The detectibility of GRBs at $z \gg 6$ has been considered in detail
by Lamb \& Reichart (2000), who find that the bright end of the
luminosity distribution would be detectable at very high
redshifts (though quantitative predictions depend substantially on
unknown details of the GRB luminosity function).  This applies
also to the X-ray and optical afterglows, for which time dilation of
the most distant afterglows helps offset the increase in luminosity
distance with redshift (Lamb \& Reichart 2000; Ciardi \& Loeb 2000).

The Lyman break will render afterglows at $z > 7$ invisible to optical
detectors, just as it does for galaxies.  But the problem here is not
so serious.  Searches for $z > 7$ galaxies 
suffer because galaxies at such high redshifts are faint, and the
required combination of large solid angles and high sensitivity to
find them is not yet practical at near-IR wavelengths.  Because GRB
afterglows outshine their host galaxies at early times, and because
X-ray detectors can determine GRB locations with accuracy comparable
to the current near-IR field of a 4m class telescope, an afterglow at
this redshift is easier to find than are the galaxies
around it.  Indeed, published near-infrared afterglow observations
(Rhoads \& Fruchter 2001) already achieve a sensitivity sufficient to
detect afterglows at $z \sim 10$ for several hours following a GRB
(cf. figures~2,3 of Lamb \& Reichart 2000).  The followup program
described in Rhoads \& Fruchter 2001 is continuing at the NASA
Infrared Telescope Facility, and we have a similar program at the
National Optical Astronomy Observatory.  The observed signature of a
$z > 7$ GRB would be a near-infrared afterglow exhibiting a Lyman
break at $\lambda_{obs} = 0.1215 (1 + z) \micron > 1 \micron$.  Such
breaks have been used to measure $z = 2.05$ for GRB 000301C (Smette et
al 2001) and to estimate $z \approx 5$ for GRB 980329 (Fruchter 1999;
see also Reichart et al 1999).  Their extension to longer wavelengths
is straightforward.  Thus, it is reasonable to expect that
$z>7$ GRBs will be detected with current technology.

The prospect of detecting gamma ray bursts at $z > 7$ opens two
possible methods of studying star formation activity at these epochs:
GRB rate evolution, which should trace star formation activity;
and pair production cutoffs in the GeV spectra of bursts, which probe
the total optical-ultraviolet background light produced by high
redshift stars.

\section{Burst Rate Evolution}
The most basic inference from the observed burst rate is that the
highest redshift where a burst has been detected $z_{max,grb}$ implies
the onset of star formation at some redshift $z_{max,*} >
z_{max,grb}$.  It is likely that in fact $z_{max,*} \approx
z_{max,grb}$:  The association of GRBs with star formation tracers
requires short progenitor lifetimes ($\ll 10^8$ years),
so the redshift difference between the first stars formed and the
earliest possible hypernovae is small.

It will be possible to go further by measuring the GRB rate as a
function of redshift, $R_{grb}(z)$, and taking it as a surrogate for
the star formation rate.  Such studies would require a
large sample (several tens) of high redshift GRBs, together with an
understanding of the selection effects that went into the sample.
This method is likely to be limited by at least two systematic
factors.  First, uncertainties in the GRB luminosity function will
introduce uncertain corrections to the inferred total GRB rate and the
inferred star formation rate, since the high redshift sample will
contain only bright bursts.  Second, evolution in the burst progenitor
population may influence the burst rate.  One plausible example is that
the GRB rate could depend on progenitor
metallicity, which is likely to be lower in the early universe.  Another is
that the stellar initial mass function (IMF) may vary,
thereby affecting the relation between GRB rate and star formation
rate, and perhaps also the shape of the GRB luminosity function.
Possible evidence for IMF variations has recently been found in at
some high redshift Lyman $\alpha$ emitting galaxies (Malhotra et
al 2001).

Overall, these complications suggest that calibration of the GRB rate
as an indicator of global star formation might be possible to within a
factor of a few.  While higher accuracies would be desirable, the
present uncertainties with more conventional star formation estimators
are not much better.  For example, rest ultraviolet continuum
measurements are corrected by a factor of $\sim 7$ for dust
absorption, and the uncertainty in this correction could easily be a
factor of two given the range of possible dust properties.

\section{Pair Production Cutoffs in GRB Spectra}
The observed spectra of gamma ray bursts sometimes extend to very high
photon energies: The EGRET experiment on the Compton Gamma Ray
Observatory detected four bursts with unbroken power law tails
extending to $\Ephot > 1 \gev$, and the Milagrito air shower
experiment has tentatively detected one burst at $\Ephot \ga 1 \tev$.
Photons with such high energies have mean free paths shorter than a
Hubble distance due to $\gamma + \gamma \rightarrow e^+ + e^-$
interactions with low energy background photons.  The threshold for
such pair production reactions is $\Ephot \ebgphot > \me^2 c^4 = (511
\kev)^2$, corresponding to the requirement that each photon have the
rest mass energy of an electron in their center of momentum frame.
(Here $\Ephot$ and $\ebgphot$ are the two photon energies measured in
an arbitrary frame, and $\Ephot \ge \ebgphot$ by convention.)  The
cross section (for a head-on collision) peaks at $\Ephot \ebgphot = 2
\me^2 c^4$ and falls asymptotically as $ 1 / (\Ephot \ebgphot)$ for
$\Ephot \ebgphot \gg 2 \me^2 c^4$.

Pair production cutoffs in the \tev\ gamma ray spectra of blazars due
to interactions with the cosmic infrared background have been
predicted (Stecker, De Jager, \& Salamon 1992; MacMinn \& Primack
1996; Madau \& Phinney 1996; Malkan \& Stecker 1998)
and observed (e.g., De Jager, Stecker, \& Salamon 1994; Konopelko et
al 1999) for several years now.  
%
The extension of the same physics to higher redshifts and lower gamma
ray energies has been explored recently by several groups (Salamon \&
Stecker 1998; Primack et al 2000; Oh 2000).

The observer frame gamma ray energy determines simultaneously the
redshift and rest frame energies of the background photons that
dominate the pair production optical depth.  At low redshifts ($z\ll
1$), the effective absorption coefficient $\alpha(\Ephot)$ increases
with $\Ephot$ and changes relatively little with redshift, so that the
relevant physics is simply $\alpha(\Ecut) d = 1$, with $d$ the
distance to the source.  However, at $z \ga 1$, redshift effects
become important: The threshold energy $\ebgphot(z) \propto 1/(1+z)$,
and the background radiation field will also evolve with redshift.
The optical depth for photons near $\Ecut$ is therefore dominated by
absorption at high redshift, unless the source redshift is so high as
to precede the creation of any substantial optical-IR background.  By
the time the photon reaches lower redshifts, the threshold for pair
creation grows so large that the density of relevant photons is
extremely low.  Oh (2000) has shown that the highest energy background
photons capable of producing optical depth $\tau \approx 1$ over a
Hubble distance have energies below the ionization threshold for
hydrogen (i.e., $\ebgphot < 13.6 \ev$), since hydrogen absorption in
stellar atmospheres, galaxies, and the intergalactic medium ensures a
strong decrement in background photon number density at $13.6 \ev$.

The most robust observable consequence of the pair creation cutoff is
the observer frame gamma ray energy $\Ecut(z)$ for which the optical
depth $\tau = 1$.  Lower pair creation optical depths ($\tau \ll 1$)
cannot be measured reliably because of our imperfect knowledge of the
intrinsic (i.e., unabsorbed) source spectrum, while at higher optical
depths ($\tau \gg 1$) the absorption reduces the flux below detection
thresholds of present or near-future instruments.  We might
measure $\tau(\Ephot)$ with reasonable accuracy over the range $1/2
\la \tau \la 2$.

Detailed predictions of $\Ecut(z)$ differ from model to model,
depending on the theoretical treatment adopted for the earliest star
formation (see Primack et al 2000; Oh 2000).  For example, the
observer frame energy where $\tau = 1$ for redshift $z=6$ is $4 \GeV
\la \Ecut(6) \la 6 \GeV$ for different models in Primack et al (2000),
and $10 \GeV \la \Ecut(6) \la 26 \GeV$ for models in Oh (2000).
Therein lies the power of this method for learning about the first
generations of stars, for these strong differences in predictions
allow the models to be distinguished with comparative ease from even a
modest data set.

Moreover, if we can observe the GeV cutoffs in spectra of a few GRBs
spread over the redshift range $6 \la z < z_{max,*}$, we can infer the
evolution of the optical-UV background radiation over the same period
with little dependence on models.  This follows because the {\it
difference} in pair creation optical depth between two bursts at
redshifts $z_1$ and $z_2$ ($z_1 < z_2$) is determined only by the
background radiation in the range $z_1 < z < z_2$.
%

\section{Discussion}
The two methods of using gamma ray bursts to probe high redshift star
formation complement each other in many ways.  GRB rate measurements
at high redshift are technically easier.  They require a GRB monitor
plus rapid multiband near-infrared followup.  Existing instrumentation
and indeed existing observational programs are already adequate for
this work.
%
Pair creation cutoffs require one additional observation, namely, a
GeV energy spectrum obtained during the GRB.  
This GeV spectrum will have to come from GLAST or a similar space
mission. 

The physical assumption behind the GRB rate evolution method is that
the bursts are associated with star formation activity.  Under this
assumption, there will be some systematic uncertainties in converting
the GRB rate to the star formation rate (see above).
%
In contrast, the pair creation cutoff method requires only that some
high redshift GRBs have GeV spectra that are sufficiently bright and
sufficiently smooth for the cutoff to be observed.
%
%
Beyond this, there is no requirement on the nature of the bursters,
which are needed only as beacons to probe the intervening background
radiation.  The physics of pair creation is then well understood
and probes the total background radiation produced by high redshift
stars.

Thus, combining the two methods of studying high redshift star
formation with GRBs may overcome the physical uncertainties of either
method alone.  Additional constraints from other techniques using
other classes of objects (galaxies observed at infrared wavelengths,
or quasars at X-ray wavelengths) will become available over the next
few years, and will again have complementary strengths and weaknesses.
By adding these to the GRB results, we can reasonably expect to
understand star formation at $z \sim 10$ as well as we understand it
at $z \sim 3$ today.


\begin{thebibliography}{foobar}
\def\aj{{\rm AJ}}                   
\def\araa{{\rm ARA\&A}}             
\def\apj{{\rm ApJ}}                 
\def\apjl{{\rm ApJ}}                
\def\apjs{{\rm ApJS}}               
\def\aap{{\rm A\&A}}                
\def\mnras{{\rm MNRAS}}             
\def\pasp{{\rm PASP}}               

\def\reference{\bibitem}

\reference{bl} Barkana, R., \& Loeb, A. 2001, {\it Physics Reports},
  in press 
\reference{bkd} Bloom, J. S., Kulkarni, S. R., \& Djorgovski, S. G. 2000,
  submitted to \aj, astro-ph/0010176 
\reference{cl} Ciardi, B., \& Loeb, A. 2000, \apj\ 540, 687
\reference{dss} De Jager, O. C., Stecker, F. W., \& Salamon, M. H. 1994, 
   \nature\ 369, 294 
\reference{f99a} Fruchter, A. S., et al 1999, \apjl\ 519, L13
\reference{f99b} Fruchter, A. S. 1999, \apjl\ 512, L1 
\reference{f01} Fruchter, A. S., et al 2001 
\reference{kksm} Konopelko, A. K., Kirk, J. G., Stecker, F. W., \&
   Mastichiadis, A. 1999, \apjl\ 518, L13  
\reference{lr} Lamb, D. Q., \& Reichart, D. E. 2000, \apj\ 536, 1
\reference{mmpr} MacMinn, D., \& Primack, J. R. 1996, {\it Space Science
   Reviews} 75, 413  
\reference{maph} Madau, P., \& Phinney, E. S. 1996, \apj\ 456, 124
\reference{m01} Malhotra, S., et al 2001, in preparation 
\reference{ms} Malkan, M. A., \& Stecker, F. W. 1998, \apj\ 496, 13
\reference{oh01} Oh, S. P. 2001, to appear in \apj, astro-ph/0005263
\reference{psbd} Primack, J. R., Somerville, R. S., Bullock, J. S., 
  \& Devriendt, J. E. G. 2000, astro-ph/0011475
\reference{r99} Reichart, D. E., et al 1999, \apj\ 517, 692
\reference{r00} Rhoads, J. E., Malhotra, S., Dey, A., Stern, D., Spinrad,
        H., \& Jannuzi, B. T. 2000, \apjl\ 545, L85
\reference{rf01} Rhoads, J. E., \& Fruchter, A. S. 2001, \apj\ 546, 117
\reference{ss98} Salamon, M. H., \& Stecker, F. W. 1998, \apj\ 493, 547
\reference{sds92} Stecker, F. W., De Jager, O. C., \& Salamon, M. H. 1992,
  \apj\ 390, L49 


\end{thebibliography}
\end{document}